\documentstyle[12pt,epsf]{article}
\begin{document}
\begin{titlepage}
\title{
\vspace{-1.5truein}
\begin{flushright}
{\normalsize
{\bf TTP 95-12\footnote[3]
{The complete preprint, including figures, is
also available via anonymous ftp at
ttpux2.physik.uni-karlsruhe.de (129.13.102.139)
as /ttp95-12/ttp95-12.ps,
or via www at
http://ttpux2.physik.uni-karlsruhe.de/cgi-bin/preprints/}
}\\
{\bf March 1995}\\
\vspace{-0.3cm}
{\bf hep-ph/9503461}
}
\end{flushright}
\vspace{0.7truein}
Polarized Heavy Quarks
\footnote[1]{ Work supported in part by
KBN grant 2P30225206 and by DFG.}
}
\author{ \\
\large Marek Je\.zabek
\thanks{$^\ast\,$Alexander-von-Humboldt Foundation fellow.} $^\ast$
\\
{\normalsize \it Institute of Nuclear Physics,
Kawiory 26a, PL-30055 Cracow, Poland}\\
{\normalsize\rm and}\\
{\normalsize \it  Institut f\"ur Theoretische Teilchenphysik,
D-76128 Karlsruhe, Germany}
}
\date{}
\maketitle
\thispagestyle{empty}
\begin{abstract}
\noindent
Polarization studies for heavy quarks can lead to important
tests of the Standard Model. Top quark pair production in
$e^+e^-$ annihilation is considered near energy threshold.
It is shown that for longitudinally polarized electrons
the produced top quarks and antiquarks are highly polarized.
Dynamical effects originating from strong interactions in the
$t-\bar t$ system can be calculated using Green function
method. Energy-angular distributions of leptons in
semileptonic decays of polarized heavy quarks
are sensitive to both the polarization of the decaying quark
and V-A structure of the weak charged current.
Some applications to $b$ quark physics at the $Z^0$ resonance
are briefly reviewed.
\end{abstract}
\vspace{1cm}
\begin{center}
{\footnotesize\it To be published in Acta Physica Polonica B\\
Proceedings of the Cracow Epiphany Conference on Heavy Quarks\\
Jan.5-6, 1995, Cracow, Poland\\
In honour of the 60th birthday of Kacper Zalewski}
\end{center}
\vspace{1cm}
\end{titlepage}
\section{Introduction}

Polarization plays a crucial role in physics of electroweak
interactions.
Starting from the fifties when parity violation was
discovered up to present days of the LEP \cite{alexan,jw}
and SLC \cite{woods} experiments,
polarized fermions in initial and final states
have been instrumental in uncovering properties of fundamental
particles and their interactions.
Quite often due to a large degree of polarization, high
accuracy can be achieved even for a relatively low number
of events. A recent spectacular example is the precise
measurement of the electroweak mixing parameter
$\sin^2\theta_{\rm w}^{eff}$ at SLC \cite{SLD}.
Nowadays many processes involving polarized leptons are
successfully employed at experimental facilities.
The situation is quite opposite for the strongly interacting
fundamental fermions. Due to confinement the quarks remain bound
inside hadrons which are strongly interacting composite systems.
Thus in general the physical quantities depend on the
polarizations of quarks in intricate manner. It is remarkable,
however, that Nature provides us with a few processes which
can be considered as sources of highly polarized top, bottom,
and charm quarks. Moreover, in these reactions the polarizations
of the heavy quarks are not much affected by strong interactions.
Some physicists believe that the third generation of quarks
is the best available window on new physics beyond the
Standard Model. Therefore, it is reasonable to expect that
future experimental studies with polarized heavy quarks will
lead to significant progress in particle physics.

In the present article some reactions are discussed which involve
polarized heavy quarks.
In Sect.2 sources of polarized top quarks are discussed.
In Sect.3 top quark pair production in $e^+e^-$ annihilation
is considered near production threshold. It is shown that the Green
function method \cite{FK,SP,JKT,Sumino1}
can be extended to the case of polarized $t$ and $\bar t$.
Some results of our recent studies \cite{HJKT,HJK} are presented.
In particular it has been demonstrated that for the longitudinally
polarized electron beam an optimally polarized sample of top quarks
can be produced.
In Sect.4 semileptonic decays of heavy quarks are discussed
including recent results on QCD corrections to these processes.
We argue that the cleanest spin analysis for the top quarks
can be obtained from their semileptonic decay channels.
In Sect.5  polarization phenomena for $b$ and $c$ quarks
produced at the $Z^0$ peak are briefly reviewed.

\section{Sources of polarized heavy quarks}

As the heaviest fermion of the Standard Model the top quark
is an exciting new window on very high mass scale physics.
There is no doubt that precise studies of top quark
production and decays will provide us with new information
about the mechanism of electroweak symmetry breaking.
The analysis of polarized top quarks and their decays
has recently attracted considerable attention;
see \cite{Kuehn3,teupitz} and references cited therein.
For non-relativistic top quarks the polarization studies
are free from hadronization ambiguities. This is due to the
short lifetime of the top quark which is shorter than
the formation time of top mesons and toponium resonances.
Therefore top decays intercept the process of hadronization
at an early stage and practically eliminate associated
non-perturbative effects.

Many processes have been proposed which can lead
to the production of polarized top quarks.
In hadronic collisions and for unpolarized beams
the polarization studies are mainly based
on the correlation between $t$ and $\bar t$ decay
products. However, single top production through $Wb$
fusion at LHC may also be a source of polarized top quarks.
An interesting reaction is top quark pair production
in $\gamma\gamma$ annihilation at a linear photon collider.
At such a machine the high energy photon beams
can be generated via Compton scattering of laser
light on  electrons accelerated in the linac.
The threshold behaviour of the reaction
$\gamma\gamma\to t\bar t$
has been reviewed in \cite{DESYWac} and the
top quark polarization in this reaction
has been recently considered in \cite{fkk}.
A linear photon collider is a very interesting project.
If built it may prove to be one of the most useful facilities
exploring the high energy frontier. However, at present
it is not clear whether the energy resolution of this accelerator
can be considerably improved. As it stands the energy resolution
limits precision of the top quark threshold studies at photon
colliders.
The most efficient and flexible reaction producing
polarized top quarks is pair production in $e^+e^-$
annihilation with longitudinally polarized electron beams.
For $e^+e^-\to t\bar t$ in the threshold region
one can study decays of polarized top quarks under
particularly convenient conditions: large event rates,
well-identified rest frame of the top quark,
and large degree of polarization. At the same time,
thanks to the spectacular success
of the polarization program at SLC \cite{woods},
the longitudinal polarization of the electron beam
will be an obvious option for a future linear collider.

\section{Top quark pair production near threshold}

\subsection{Green function method}
The top quark is a short--lived particle. For the top mass
$m_t$ in the range 160--190 GeV its width
$\Gamma_t$ increases with $m_t$ from 1 to 2 GeV.
Thus $\Gamma_t$  by far exceeds
the tiny ($\sim$ 1 MeV) hyperfine splitting for toponia and
open top hadrons,
the hadronization scale of about 200 MeV,
and even the energy splitting between $1S$ and $2S$ $t\bar t$
resonances. On one side this is an advantage because
long-distance phenomena related to confinement are
less important for top quarks \cite{KuAct,BDKKZ}.
In particular depolarization due to hadronization is
practically absent.
On the other side the amount of information available from
the threshold region is significantly reduced. Toponium
resonances including the $1S$ state overlap each other.
As a consequence the cross section for $t\bar t$ pair production
near energy threshold has a rather simple and smooth shape.

\begin{figure}[htb]
\epsfxsize=5.5in
\leavevmode
\epsffile[120 550 580 650]{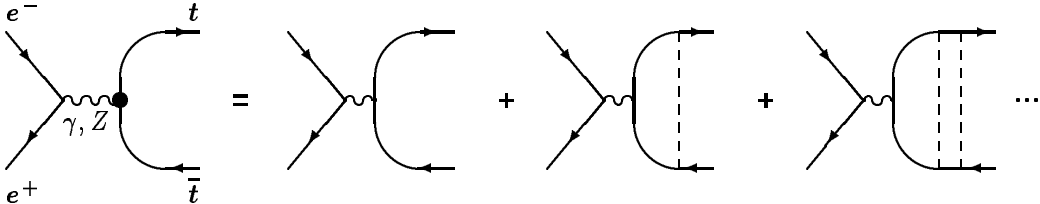}
\vskip-0.2cm
\caption{Dominant contributions to the process
$e^+e^-\to t\bar t$ near threshold.}
\label{fig-ladder}
\end{figure}
The quantitative theoretical study of the threshold region is
a complicated problem. The excitation curve
$\sigma(e^+e^-\rightarrow t\bar t\,)$
depends on $m_t$ and $\Gamma_t$. In addition
it is drastically affected by strong interactions.
A few GeV below and above the nominal threshold
$\sqrt{s}\,=\,2 m_t$ a multitude of overlapping
$S$ wave resonances  is excited.
One might think that a reasonably accurate description can
be obtained by performing a sum over these resonances.
However, it has been shown \cite{Kwong} that one has to include
so many resonances that such an approach is practically useless.
Perturbative approach is also non-trivial in the threshold region.
In seminal papers \cite{FK} Fadin and Khoze have demonstrated that
for non-relativistic $t$ and $\bar t$
the dominant contribution to the amplitude is given
by the sum of the ladder diagrams depicted in Fig.\ref{fig-ladder}.
The dashed lines denote the instantaneous parts of the gluon
propagators which in the Coulomb gauge read
\begin{equation}
D^{\mu\nu}\left( q^2 \right)\; \sim\; \delta^{\mu 0}\delta^{\nu 0}\,
V\left(\,{\bf q}\,\right)
\end{equation}
where ${\bf q}$ denotes the three-momentum transfer and
$V\left(\,{\bf q}\,\right)$ is the chromostatic potential
in the momentum space. We have also neglected contributions
of space-space components $D^{ij}$ which are suppressed
by factors of order $\beta^2$.
The diagram with $n$ exchanges
gives the contribution of order $(\alpha_s/\beta)^n$ where
$\alpha_s$ is the strong coupling constant and $\beta$ denotes
the velocity of the top quark in the center-of-mass frame.
In the threshold region $\beta \sim \alpha_s$ and
all the contributions are of the same order. In \cite{FK} it has
been also shown that the sum of the terms in Fig.\ref{fig-ladder}
can be expressed through the Green function of the $t-\bar t$
system. The effects of the top quark width have been
incorporated through the complex energy $E\,+\,i\Gamma_t$,
where
$$E=\sqrt{s} - 2m_t$$
is the non-relativistic energy of the
system. Finally, Fadin and Khoze \cite{FK} have calculated
analytically the Green function for the Coulomb chromostatic
interaction between $t$ and $\bar t$. They have pointed out that
the excitation curve $\sigma(e^+e^-\rightarrow t\bar t\,)$
allows a precise determination of $m_t$ as well as of other
quantities such as $\Gamma_t$ and $\alpha_s$.
Strassler and Peskin \cite{SP} have obtained similar results
using a numerical approach and a more realistic QCD potential.
The idea \cite{FK,SP} to use the Green function
instead of summing over overlapping resonances has been also
applied in numerical calculations of differential cross sections.
Independent approaches have been developed for solving Schr\"odinger
equation in position space \cite{Sumino1} and Lippmann-Schwinger
equation in momentum space \cite{JKT,JT}.
The results of these two methods agree very well \cite{Martinez}.
One of the most important future applications will be
the determination of $m_t$ and $\alpha_s$. More detailed discussions
can be found in the original papers and in the recent reviews
\cite{Kuehn3,teupitz,DESYWac,Khoze}.

\subsection{Vertices and Lippmann-Schwinger equations}
It has been already mentioned that the hyperfine splitting for
$t-\bar t$ system is much smaller than its lifetime. This implies
that the polarizations of $t$ and $\bar t$ are only weakly
affected by QCD interactions between these quarks. It is natural,
therefore, to consider the production of $t$ quark
(and $\bar t$ antiquark) of given
polarization. For the sake of simplicity we confine our discussion
to the case of top quark polarization. Two-particle spin correlations
for the $t-\bar t$ system will be discussed elsewhere \cite{HJK}.
In close analogy to the unpolarized case, c.f. Fig.\ref{fig-ladder},
we consider $e^+e^-$ annihilation into $t\bar t$ pair.
The four-momentum of the top quark is denoted by
$p_+$ and its spin four-vector by $s_+$.
The antiquark $\bar t$ carries the four-momentum $p_-$.
The electron and the positron are relativistic and their
masses can be neglected.
Let $k_\pm$ denote the four-momenta
of $e^\pm$  ($k_\pm^2 = 0$),
$$Q\> =\> k_-\, +\, k_+ \> = \> \left(\sqrt{s},0,0,0\right)$$
and
$$K\> =\> k_-\, -\, k_+ \> = \> \left(0,0,0,\sqrt{s}\right)$$
The matrix element squared for $e^+e^-\to t\bar t$ can be written
as a contraction of the leptonic and hadronic tensors
\begin{equation}
\left|{\cal M}\right|^2\; \sim \; L^{\alpha\beta}\, H_{\alpha\beta}
\label{eq-M2}
\end{equation}
It is evident from Fig.\ref{fig-ladder} that the leptonic tensor
$L^{\alpha\beta}$ is well described by the Born expression
whereas the hadronic tensor $H_{\alpha\beta}$
is given by a complicated sum of ladder diagrams. Let $J_z$
denote the component of the total angular momentum in the
direction of $e^-$. Then
$$L^{\alpha\beta} = 0 \quad {\rm for} \quad J_z=0$$
whereas for for $J_z = \pm 1$
\begin{eqnarray}
L^{\alpha\beta}_{VV}\; =\; L^{\alpha\beta}_{AA} &=&
L^{\alpha\beta}_{s}\, +\, J_z\, L^{\alpha\beta}_{a}
\nonumber\\
L^{\alpha\beta}_{VA}\; =\; L^{\alpha\beta}_{AV} &=&
J_z\, L^{\alpha\beta}_{s}\, +\,  L^{\alpha\beta}_{a}
\label{eq-lepton}
\end{eqnarray}
The subscripts $A$ and $V$ denote the contributions of
the vector and axial-vector leptonic currents, and
\begin{eqnarray}
L^{\alpha\beta}_{s} &\sim& s g^{\alpha\beta} - Q^\alpha Q^\beta
+ K^\alpha K^\beta
\nonumber\\
L^{\alpha\beta}_{a} &\sim& \varepsilon^{\alpha\beta\lambda\mu}
Q_\lambda K_\mu
\end{eqnarray}
It follows from eq.(\ref{eq-lepton}) that for longitudinally
polarized electrons and positrons the total annihilation
cross section is proportional to
$$  1 - P_{e^+}P_{e^-}$$
where $P_{{e^+}}$ and $P_{{e^-}}$ denote the polarizations
of $e^+$ and $e^-$, with respect to the directions of $e^+$
and $e^-$ beams, respectively. Furthermore,
the polarization of the top quark depends only on the variable
\begin{equation}
\chi = {P_{e^+}-P_{e^-}\over 1 - P_{e^+}P_{e^-}}
\end{equation}
It is conceivable that for a future linear $e^+e^-$ collider
$P_{e^+}=0$, $P_{e^-}\ne 0$ and  $\chi = -P_{e^-}$.
Another interesting observation is that only the space-like
components $H^{ij}$ of the hadronic tensor can contribute
to the differential cross section. (In fact only the transverse
components $i,j=1,2$ give non-zero contributions.) Thus in the
following discussion we consider only the components $H^{ij}$ of
the hadronic tensor.

\begin{figure}
\epsfxsize=5.5in
\leavevmode
\epsffile[90 530 600 660]{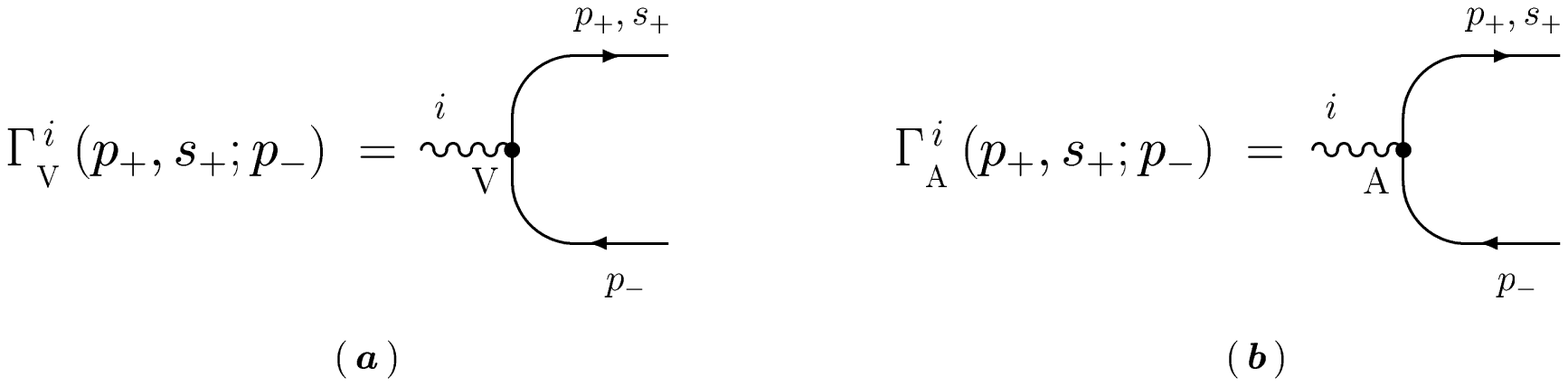}
\caption{Effective vertices describing the couplings of
a) the vector and b) the axial-vector current to
the top quark of four-momentum $p_+$ and spin
four-vector $s_+$ and  the antiquark $\bar t$ of four-momentum
$ p_-$.}
\label{fig-Vert}
\end{figure}
\begin{figure}
\epsfxsize=5.5in
\leavevmode
\epsffile[130 550 600 650]{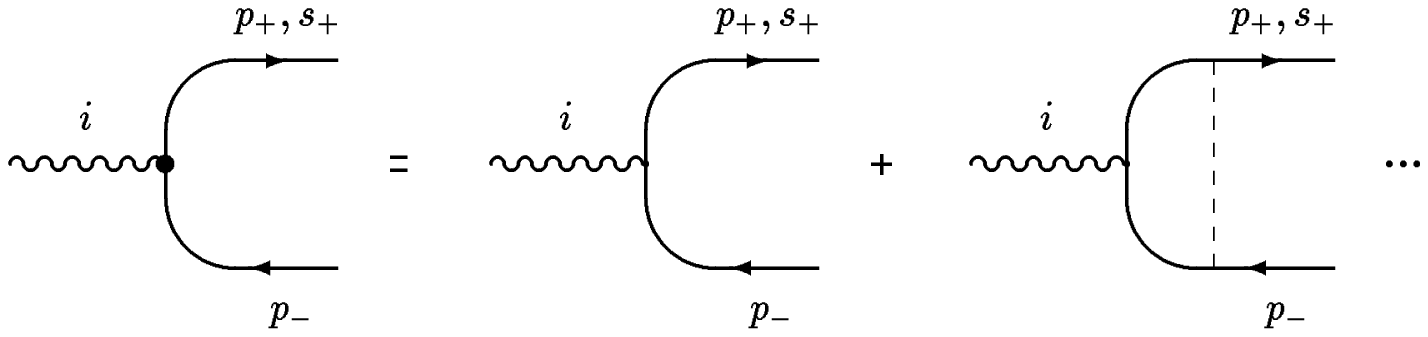}
\vskip-0.2cm
\caption{Definition of the effective vertices.}
\label{fig-Vertd}
\end{figure}
In the center-of-mass frame
the velocity of the top quark is small
($\beta = |{\bf p}|/m_t \> \ll \> 1$) and we can use
the non-relativistic approximation for $t$ and $\bar t$.
The spin four-vector is
\begin{equation}
s_+^\mu = (0,{\bf s}_+) + {\cal O}(\beta)
\end{equation}
Thus, up to terms of order $\beta^2$ the spin three-vector
${\bf s}_+$ is the same as in the top quark rest frame. We define
effective vertices  $\Gamma^i_V$ and $\Gamma^i_A$
describing the couplings of
the vector ($V$) and the axial-vector currents ($A$) to
the top quark of four-momentum $p_+$ and spin
four-vector $s_+$ and  the antiquark $\bar t$ of four-momentum
$ p_-$, see Fig.\ref{fig-Vert}.
Each of these vertices is an infinite sum
of ladder diagrams corresponding to instantaneous Coulomb-like
exchanges of gluons between $t$ and $\bar t$, see Fig.\ref{fig-Vertd}.
The space-like components of the hadronic tensor $H^{ij}$ can be
expressed through the effective vertices
\begin{equation}
H^{ij} \> \sim\> \sum_{a,b}\, Tr\,
\left[ \Gamma^i_a\widetilde{\Gamma^j_b}\right]
\end{equation}
where $a,b\, =\, V,A$ and
$\widetilde{O} = \gamma^0 O^\dagger \gamma^0 $. Let us define now
the projection operators
\begin{equation}
\Lambda_\pm \> = \> {\textstyle{1\over 2}}
\left(1 \pm \gamma^0 \right)
\end{equation}
We can split any operator $O$ into two pieces $(O)_\pm$:
\begin{eqnarray}
(O)_+&=& \Lambda_+ O \Lambda_+ \,+\, \Lambda_- O \Lambda_-
\nonumber\\
(O)_-&=& \Lambda_+ O \Lambda_- \,+\, \Lambda_- O \Lambda_+
\end{eqnarray}
which we call even and odd parts of $O$ respectively.
It can be shown that
\begin{equation}
\Gamma^i_V = \left( \Gamma^i_V \right)_- + {\cal O}(\beta)
\qquad {\rm and} \qquad
\Gamma^i_A = {\cal O}(\beta)
\nonumber
\end{equation}
Any product of an odd and an even operator is traceless, so
up to terms of order $\beta^2$ only odd parts of $\Gamma^i_a$
can contribute to $H^{ij}$.
\begin{equation}
H^{ij} \> \sim\> \sum_{a,b}\, Tr\,
\left[ (\Gamma^i_a)_{_-}(\widetilde{\Gamma^j_b})_{_-}\right]\,
+\, {\cal O}\left(\beta^2\right)
\end{equation}
Furthermore, with the same accuracy the effective vertices can
be expressed through two scalar functions:
${\cal K}_V\left( p,E\right)$ and
${\cal K}_A\left( p,E\right)$
where $p = |{\bf p}|$.
It follows from
Fig.\ref{fig-Vertd} that
\begin{eqnarray}
\left(\, \Gamma^j_V(p_+,s_+;p_-)\, \right)_- &=&
\Lambda_+ \Sigma_+ \gamma^j \Lambda_-\>
{\cal K}_V\left( p,E\right)
\label{eq-GamV}
\\
\left(\, \Gamma^j_A(p_+,s_+;p_-)\, \right)_- &=&
{i\over m_t}\,
\Lambda_+ \Sigma_+ ({\bf\vec\gamma\times p})^j
\Lambda_-\>
{\cal K}_A\left( p,E\right)
\label{eq-GamA}
\end{eqnarray}
where
\begin{equation}
\Sigma_+\> = \> {\textstyle{1\over 2}}
\left( 1 + {\bf s_{\rm +}\cdot \Sigma} \right)
\end{equation}
and  $\Sigma^i\> =\> \gamma_5\,\gamma^0\,\gamma^i$
is the Dirac spin operator.
\begin{figure}
\epsfxsize=5.5in
\leavevmode
\epsffile[130 550 530 650]{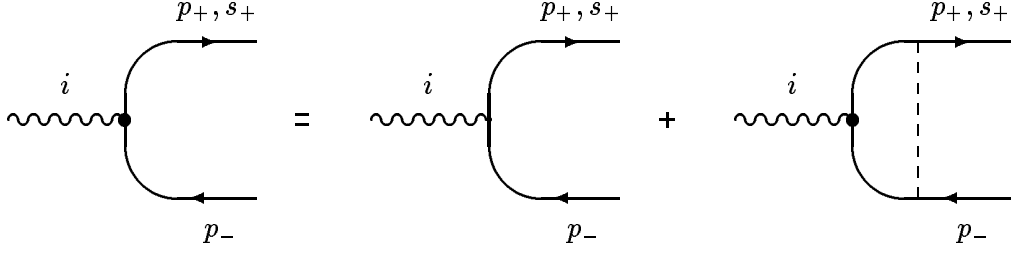}
\vskip-0.2cm
\caption{Lippmann--Schwinger equation for the
effective vertices.}
\label{fig-LSeq}
\end{figure}
The series defining the
effective vertices, see Fig.\ref{fig-Vertd},
can be formally summed. In this way the equation depicted
in Fig.\ref{fig-LSeq} is derived. Neglecting the corrections of
order $\beta^2$ one obtains the following integral equations for
the functions
${\cal K}_V\left( p,E\right)$
and
${\cal K}_A\left( p,E\right)$
\begin{eqnarray}
\hskip-15pt
{\cal K}_V\left( p,E\right) &=&
1 + \int {d^3 q\over(2\pi)^3}\, V\left( {\bf p - q} \right)
G_0(q,E)\, {\cal K}_V\left( q,E\right)
\label{eq-LS0}
\\
\hskip-15pt
{\cal K}_A\left( p,E\right) &=&
1 + \int {d^3 q\over(2\pi)^3}\,{{\bf p\cdot q}\over p^2}\,
V\left(\, {\bf p - q} \right)
G_0(q,E)\, {\cal K}_A\left( q,E\right)
\label{eq-LS1}
\end{eqnarray}
where $G_0(p,E)$
is the free Green function for the $t-\bar t$ system\footnote{
It is consistent to neglect the momentum dependence
of the width for the non-relativistic $t-\bar t$ system because
the corresponding corrections are of order $\beta^2$.
Recent measurements by CDF collaboration\cite{CDF} imply
$m_t = 176\pm 8(stat.) \pm 10(sys.)$~GeV and the analysis of
$D\O$ collaboration\cite{D0} gives
$199^{+19}_{-21}(stat.)\pm 22(sys.)$~GeV.
It has been shown that the corrections
due to momentum dependent width cancel to large extent
and are quite small for $m_t\>\sim\> 180$~GeV \cite{JT,KuM}.}
\begin{eqnarray}
G_0(p,E) \hskip-5pt
&=& \hskip-5pt
{-1\over 2\pi i}\int{dp^0\over
\left(p^0 - {{\bf p}^2\over 2m_t} + {\rm i}{\Gamma_t\over 2}\right)
\left(E-p^0 - {{\bf p}^2\over 2m_t} + {\rm i}{\Gamma_t\over 2}\right)}
\>=\> {1\over E - {p^2\over m_t} +
{\rm i}\Gamma_t}
\nonumber\\  &&
\end{eqnarray}
It can be shown that the function
\begin{equation}
G(p,E) \> =\> G_0(p,E)\,
{\cal K}_V\left( p,E\right)
\end{equation}
is the $S$ wave Green function \cite{FK,SP}.
It solves the following Lippmann-Schwinger equation
\begin{equation}
G\left( p,E\right) =
G_0(p,E)\> + \> G_0(p,E)\,
\int {d^3 q\over(2\pi)^3}\, V\left({\bf p - q} \right)\,
G(q,E)
\end{equation}
which follows trivially from eq.(\ref{eq-LS0}). The function
\begin{equation}
F(p,E) \> =\> G_0(p,E)\,
{\cal K}_A\left( p,E\right)
\end{equation}
is related to the $P$ wave Green function \cite{MS}.
The Lippmann-Schwinger equation for $F(p,E)$
follows from eq.(\ref{eq-LS1}):
\begin{equation}
F\left( p,E\right) =
G_0(p,E)\> + \> G_0(p,E)\,
\int {d^3 q\over(2\pi)^3}\,
{{\bf p\cdot q}\over p^2}\,
V\left({\bf p - q} \right)\,
F(q,E)
\end{equation}

A remarkable feature of the odd parts of the effective vertices
$\Gamma^j_V(p_+,s_+;p_-)$
and
$\Gamma^j_A(p_+,s_+;p_-)$
is that in non-relativistic approximation their spinor structures
are not changed by chromostatic interactions,
see eqs.(\ref{eq-GamV}) and (\ref{eq-GamA}).
When these interactions are switched off (i.e. $V=0$)
the sums of the ladder diagrams which define the effective
vertices reduce to single diagrams with no gluon exchanges,
the vertex functions
${\cal K}_V\left( p,E\right)$ and
${\cal K}_A\left( p,E\right)$ become equal one,
and the spinor structures remain the same.
This means that from a practical point of view the calculations of
the matrix element squared (\ref{eq-M2}) including chromostatic
interactions between $t$ and $\bar t$ can be reduced to the evaluation
of Born contributions. The only difference is that the vector
and axial-vector couplings $g_v$ and $g_a$ of the quark current
to photon and $Z^0$ are modified:
\begin{eqnarray}
g_v\; \to \; \tilde g_v = g_v(4m_t^2)\,
{\textstyle \left(1 - {8\alpha_s\over 3\pi}\right)}\,
{\cal K}_V\left( p,E\right)
\label{eq-gV}\\
g_a\; \to \; \tilde g_a = g_a(4m_t^2)\,
{\textstyle \left(1 - {4\alpha_s\over 3\pi}\right)}\,
{\cal K}_A\left( p,E\right)
\label{eq-gA}
\end{eqnarray}
The prescription given in eqs.(\ref{eq-gV}) and (\ref{eq-gA})
includes not only chromostatic interactions but also two
other important effects: the scale dependence of the running
coupling constants $g_{v,a}(4m_t^2)$
and the factors
$\left(1 - 8\alpha_s/ 3\pi\right)$
and
$\left(1 - 4\alpha_s/ 3\pi\right)$ which arise from
loop integrations over the relativistic
region (contributions of hard transverse gluons);
see \cite{SP} and \cite{kz}.

\subsection{Cross sections}
We are ready now to describe the process which consists
of the emission of a $t-\bar t$ system by a virtual photon
or $Z^0$ and its subsequent propagation
and decay into $\bar t W^+b$
(or $t W^-\bar b$). This is just the most difficult
part of the calculation for which perturbative (in $\alpha_s$)
approach is not adequate. After the decay the
time evolution of the system
is governed by the free motion of $W^+$
and chromodynamical interactions in the $\bar t - b$ system.
(If $\bar t$ decays first one considers the analogous time evolution
for $W^-$ and $t-\bar b$). In this period one of the strongly
interacting fermions is relativistic.
In contrast to the case of the $t-\bar t$ system
the summation over ladder diagrams
is not necessary because a diagram with $n$ exchanged gluons
is suppressed by $\alpha_s^n$. In other words this part of the
time evolution can be described in ordinary perturbative approach.
Finally the $W^-\bar t b$ system decays into $W^-W^+\bar b b$.
The amplitudes ${\cal F}_{1,2}$ describing the two decay
sequences in $t-\bar t$ rest frame
\begin{eqnarray}
{\cal F}_1\; :\qquad\bar t t\to \bar t W^+ b \to W^-W^+ \bar b b
\nonumber\\
{\cal F}_2\; :\qquad \bar t t\to W^- \bar b t \to W^-W^+ \bar b b
\nonumber
\end{eqnarray}
have to be added coherently. The theoretical description becomes
even more complicated when $W$ bosons are treated as unstable
particles. In such a case we have six different decay sequences.
Furthermore one or two of $W$ bosons can decay into quarks
whose interactions with $b$ and/or $\bar b$ can be also
important in some regions of phase space. These are the so-called
{\it cross talking} or {\it interconnection} effects\cite{Khoze}.
Even more important are effects of gluon radiation off
$\bar t-b$ and $t-\bar b$ systems\cite{KOS,KhS}. All these
phenomena have to be included into a complete theoretical
analysis of $t\bar t$ production near threshold. However,
it is likely that these refinements will not drastically change
the results for inclusive cross sections which we consider in the
following.  In fact, we assume that the contributions
of the interference terms cancel. This assumption can be easily
justified when QCD interactions in $\bar tW^+b$ and $W^-\bar b t$
systems are neglected. Let $p^0$ denote the energy of $t$ which
for non-interacting system is equal to the total energy $W^+ b$.
Overall energy conservation implies that the energy of $W^-\bar b$
system (i.e. of $\bar t$) is equal to $\sqrt{s} - p^0$.
The product of propagators for $t$ and $\bar t$ can be
written as a sum of two terms corresponding to the two different
sequences of decays
\begin{equation}
G^t_0(p^0,{\bf p})\, G^{\bar t}_0(\sqrt{s}-p^0,{\bf -p}) \; = \;
G_0(p,E)\,
\left[\, G^{\bar t}_0(\sqrt{s}-p^0,{\bf -p})\, +
G^t_0(p^0,{\bf p})\,\right]
\end{equation}
where
\begin{equation}
G^t_0(p^0,{\bf p})\> =\> G^{\bar t}_0(p^0,{\bf p})\> =\>
{1\over  p^0 - m_t - {{\bf p}^2\over 2m_t} +
{\rm i}{\Gamma_t\over 2} }
\end{equation}
Evidently the Fourier transform of the first term decribes
time evolution of the $t-\bar t$ system up to the moment
of $t$ decay and the subsequent time evolution of $\bar t$.
Thus this term corresponds to the situation when $t$ decays
before $\bar t$. The other term corresponds to the case
when $\bar t$ decays before $t$.
Neglecting some common factors
we obtain the following expressions for the
amplitudes ${\cal F}_1$ and ${\cal F}_2$:
\begin{eqnarray}
{\cal F}_1 & \sim &
G_0(p,E)\, G^t_0(p^0,{\bf p})
\\
{\cal F}_2 & \sim &
G_0(p,E)\,
G^{\bar t}_0(\sqrt{s}-p^0,{\bf -p})
\end{eqnarray}
and, consequently,
\begin{equation}
\int dp^0\, \left|\, {\cal F}_1 + {\cal F}_2\, \right|^2 \; = \;
\int dp^0\,\left(\, \left|\, {\cal F}_1\, \right|^2 +
\left|\, {\cal F}_2\, \right|^2 \, \right)
\label{eq-F1-2}
\end{equation}

Let us consider now the effects of gluon emission and QCD
interactions in $t\bar b$ and $b\bar t$ systems. As already
explained the effects of rescattering for $t\bar b$
and $b\bar t$ can be included as order $\alpha_s$ perturbations.
Other effects like real gluon emission and $tbW$ vertex corrections
decrease the top quark width $\Gamma_t$ by a correction
of order $\alpha_s$ \cite{JK1,TopW}\footnote{The complete formula
including $b$ quark mass and $W$ width
has been obtained~\cite{JK1} for a free top quark.
It is known \cite{Khoze,KOS} that interference affects the
gluon spectrum in $t\bar t$ pair production
for $E_g\sim \Gamma_t$.
However, $\delta\Gamma_t$ is infrared
finite, so the relative correction to the width
due to these effects should be only of order $\alpha_s\Gamma_t/m_t$
and can be neglected.}
\begin{equation}
\delta\Gamma_t\; \approx\; - \Gamma_0\,
{\textstyle {2\alpha_s\over 3\pi}
\left( {2\pi^2\over 3} - {5\over 2} - 3y \right) }
\end{equation}
where $y= m_{\rm w}^2/m_t^2$ and
\begin{equation}
\Gamma_0 =  {G_F\, m_t^3\over 8\sqrt{2}\pi} (1-y)^2 (1+2y)
\end{equation}
The resulting reduction of the width of about\footnote{
The numerical
value of $\delta\Gamma_t$ depends on the choice of the scale $\mu$
for running $\alpha_s(\mu)$. A widespread belief is that $\mu\sim m_t$
is a reasonable value. However, arguments in favour of a much lower
scale $\mu=0.12\, m_t$ have been also given
in the literature\cite{SmV}.} 10\% changes significantly the time
evolution of the $t\bar t$ system and affects the results
for the total cross section in the threshold region.
The rescattering corrections change the wave function
$\psi$ of $t\bar b$ (or $b\bar t$) to
$\psi^\prime = \psi + \alpha_s\delta\psi$
where up to corrections of order $\Gamma_t/m_t$
the functions $\delta\psi$ and $\psi$ are orthogonal
$\int \psi\delta\psi^* = 0 $
as a consequence of unitary time evolution.
Thus there is no ${\cal O}(\alpha_s)$ contribution to the total cross
section from the rescattering corrections. This fact, which was
first observed long ago for the electromagnetic corrections to
the lifetime of the muon bound in nuclei~\cite{Uber}, has been
recently demonstrated by explicit calculations also
for the top quark pair production~\cite{MeYa,Sumi3}.
Corrections to the differential distributions
($\sim\alpha_s\Re(\psi\delta\psi^*)\,$) have been calculated
in \cite{Sumi3}. The results fully confirm intuitive expectations
that rescattering in $b-\bar t$ system leads to reduction of
intrinsic momentum of $\bar t$ in the overall center-of-mass frame.
In this frame the $b$ quark is slowed down by the chromostatic
field of $\bar t$. Since $W^+$ is colorless it propagates as a free
particle. In consequence the total three-momentum of the $bW^+$
system decreases, which through momentum conservation implies
reduction of the intrinsic momentum for $\bar t$.
In the following discussion we neglect this correction
to the top quark momentum distributions because it only
weakly affects polarizations.

Throughout this article all corrections of order $\beta^2$
are systematically neglected. However, close to threshold
the dependence of cross sections on the width is enhanced, so
a few remarks on order $\alpha_s^2$ corrections to the width
of the $t-\bar t$ system are in order here. It has been pointed
out in Ref.\cite{Sumino1} that effects of phase space suppression
are important and cannot be neglected in quantitative studies.
As an example of the phase space suppression effects
let us consider $t\bar t$ for negative non-relativistic
energy $E\sim -\alpha_s^2 m_t$ and assume that $t$ decays first.
The propagator function $G^{\bar t}_0(\sqrt{s}-p^0,{\bf -p})$
is peaked for the energy of $\bar t$ close
to the classical value. Taking into account the kinetic energies
which according to virial theorem are of order $|E|$
one concludes that the invariant mass of the $W^+b$ system
is likely to be a few percent smaller than $m_t$. This implies
an even larger reduction of the decay rate. However, it has
been conjectured\cite{JT} and proven\cite{KuM} that order
$\alpha_s^2$ rescattering corrections to the total cross section
nearly cancel the negative contributions of phase space suppression.
The remainder can be
interpreted as due to time dilatation factors
for $t$ and $\bar t$ in the center-of-mass frame. Its effect on
the total cross section is quite small and will be neglected
in the following discussion. This implies that the volumes
of the phase spaces for the $W^+b$ and $W^-\bar b$ systems
can be considered equal and proportional to $\Gamma_t$. In
this way ${\cal O}(\alpha_s)$ corrections to the top width
are automatically included. Integration over $p^0$ in
eq.(\ref{eq-F1-2}) can be easily performed
\begin{equation}
\int\, dp^0\,\Gamma_t\, |\, G^t_0(p^0,{\bf p})\, |^2\; =\;
\int\, dp^0\, \Gamma_t\,
|\, G^{\bar t}_0(\sqrt{s} - p^0,{\bf -p})\, |^2\; =\;
2\pi
\end{equation}
Inclusive differential cross section
for the top quark production reads
\begin{eqnarray}
{d\sigma({\bf p,s_+})\over dp\, d\Omega}\> = \>
{1\over 2}\, \left(\, 1 +  2\,{\cal A}_{FB} \cos\theta +
\vec{\cal P}\cdot{\bf s}_+\, \right)\,
{d\sigma\over dp}
\label{eq-difdist}
\end{eqnarray}
where $\vec{\cal P}$ characterizes the final polarization
of the top quark,
\begin{equation}
{d\sigma\over dp} \; =\;
 \int d\Omega\> \sum_{\pm {\bf s_+}}\,
{d\sigma({\bf p,s_+})\over dp\, d\Omega}
\end{equation}
denotes its momentum distribution,
and ${\cal A}_{FB}$ is the forward-backward asymmetry.
Collecting all the factors we obtain the following expressions:
\begin{equation}
{d\sigma\over dp} \; =\;
{12\, \alpha^2(4m_t^2) \over s\, m_t^2}\,
{\textstyle \left( 1 -{8\alpha_s\over 3\pi}\right)^2}\>
\left(\,1 - P_{e^+} P_{e^-}\, \right)\>
\left( a_1\, + \chi a_2 \right)\>
\Gamma_t\, |\, p\,G(p,E)\,|^2
\end{equation}
\begin{equation}
{\cal A}_{FB}(p,E,\chi) \,=\,
{a_3 + \chi a_4 \over 2(a_1 + \chi a_2)}
\,\varphi_R(p,E)
\label{eq-AFB}
\end{equation}
The coefficients $a_1,\dots,a_4$ are given in Ref.\cite{HJKT}.
They depend on the electroweak couplings of
$\gamma$ and $Z^0$ to the electron and top quark.
The function $\varphi_R(p,E)$ is defined as the real part of
\begin{equation}
\varphi(p,E)\, =\,  {
\left( 1- {4\alpha_s\over 3\pi} \right)\> p\, F^*(p,E)
\over
\left( 1- {8\alpha_s\over 3\pi} \right)
m_t\, G^*(p,E) }
\label{eq-phi}
\end{equation}
Eq.(\ref{eq-AFB}) has been first obtained in \cite{MS} for $\chi=0$.

The Lippmann-Schwinger equations (\ref{eq-LS0}) and (\ref{eq-LS1})
can be solved numerically using the method decribed in \cite{JKT,JT}.
$S$ wave dominates the total cross section. Neglecting terms of
order $\beta^2$ one obtains the following form of the optical theorem
\begin{equation}
\int_0^\infty dp\, p^2\, |G(p,E)\,|^2 \> =\> -\,
\int_0^\infty dp\, p^2\, \Im\,G(p,E)
\end{equation}
which can be used as a cross check of numerical calculations.

\begin{figure}
\epsfxsize=5.5in
\leavevmode
\epsffile[30 310 535 510]{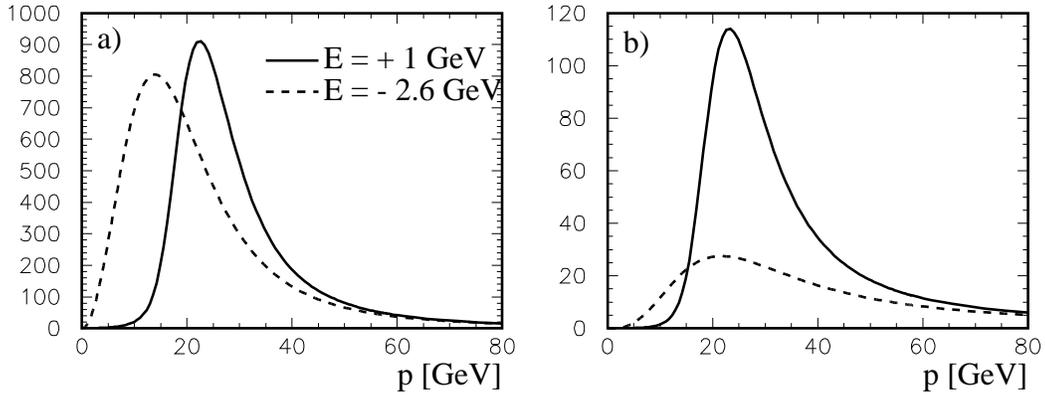}
\vskip-0.2cm
\caption{Top quark momentum and angular distributions for
$E=$ 1 and -2.6 GeV -- solid/dashed lines, $m_t = 174$ GeV
and $\alpha_s(m_Z)=\, 0.12$: a) ${\cal D}_{S-S}(p,E)$ and
b) ${\cal D}_{S-P}(p,E)$.}
\label{fig-Dss-sp}
\end{figure}
It follows from eq.(\ref{eq-difdist}) that for the unpolarized
electron and positron beams the momentum-angular distribution
of the top quark is governed by the two functions
\begin{eqnarray}
{\cal D}_{S-S}(p,E) &=& p^2 \left|\, G(p,E)\,\right|^2\\
{\cal D}_{S-P}(p,E) &=& p^3\,
\Re\left( \,G(p,E)\,F^*(p,E)\,\right)\,/m_t
\end{eqnarray}
which are shown in Fig.\ref{fig-Dss-sp}.

\subsection{Polarizations \protect\cite{HJKT,HJK}}
The polarization state of the top quark is given by the three-vector
$\vec{\cal P}$. In an orthogonal system of coordinates we can choose
any of the axes to quantize the projection of the top quark spin.
This choice determines the form of the four-vector $s_+$ whose
space component ${\bf s}_+$ is directed along the quantization axis
and the time component is fixed by the requirement $s_+\, p_+=0$.
Then the projection of the polarization three-vector $\vec{\cal P}$
on the quantization axis is obtained. It is equal to the ratio of the
difference and the sum
of the cross sections for the spin four vectors $s_+$ and $-s_+$.
Our righthanded system of coordinates is defined through the
triplet of orthogonal unit vectors: $\hat n_{\bot}$, $\hat n_{_N}$
and $\hat n_{^\|}$ where
$\hat n_{^\|}$ points in the direction of the $e^-$ beam,
$\hat n_{_N}\sim \vec p_{e^-}\times \vec p_t$  is normal
to the production plane and
$\hat n_{\bot}=\hat n_{_N}\times\hat n_{^\|}$.
This system defines the three projections of the polarization
vector $\vec{\cal P}$.
The definition of ${\cal P}_{^\|}$, ${\cal P}_{\bot}$
and ${\cal P}_N$ with respect
to the beam direction is convenient for the treatment
close to threshold and differs from the definition
of \cite{krz} where the quantities have been defined with respect
to the direction of flight of the top quark.
The angle $\vartheta$ denotes the angle between $\hat n_{^\|}$
and the three-momentum $\bf p$ of the top quark.
As already stated in the preceding subsection we neglect
rescattering corrections which will be discussed elsewhere.
Retaining only the terms up to
${\cal O}(\beta)$ one derives the following expressions for
the components of the polarization vector, as functions
of $E$, $p$, $\vartheta$ and $\chi$:
\begin{eqnarray}
{\cal P}_{^\|}(p,E,\vartheta,\chi) &=& C^0_{^\|}(\chi)
+ C^1_{^\|}(\chi)\,\varphi_R(p,E)\,\cos\vartheta
\label{eq-Ppar}\\
{\cal P}_\bot(p,E,\vartheta,\chi) &=&C_\bot(\chi)\,
\varphi_R(p,E)\, \sin\vartheta
\label{eq-Pperp}\\
{\cal P}_N(p,E,\vartheta,\chi) &=&C_N(\chi)\,
\varphi_I(p,E)\, \sin\vartheta
\label{eq-Pnorm}
\end{eqnarray}
where $\varphi_R(p,E)$ and $\varphi_I(p,E)$ denote the real
and imaginary parts of the function $\varphi(p,E)$
defined in eq.(\ref{eq-phi})
\begin{equation}
\varphi_R(p,E)\> =\> \Re\varphi(p,E)\ , \qquad\qquad
\varphi_I(p,E)\> =\> \Im\varphi(p,E)
\end{equation}
\begin{figure}
\epsfxsize=5.5in
\leavevmode
\epsffile[30 310 535 510]{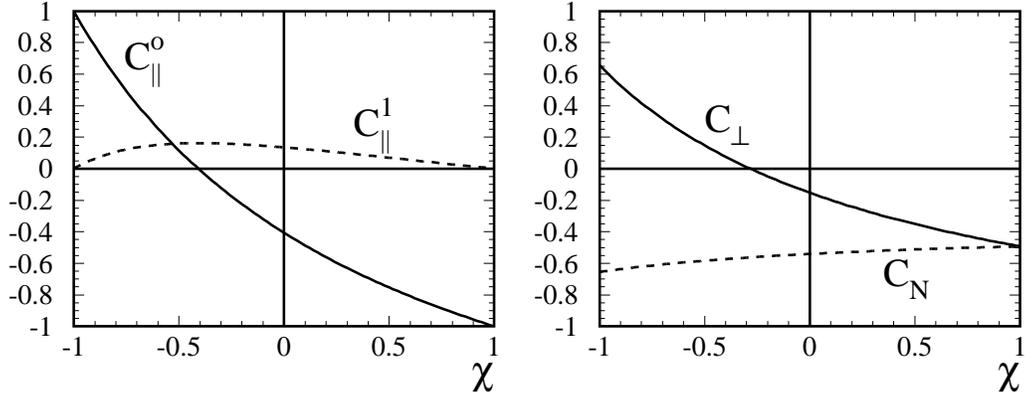}
\vskip-0.2cm
\caption{Coefficient functions: a) $C^0_{^\|}(\chi)$ -- solid line
and $C^1_{^\|}(\chi)$ -- dashed line, b) $C_\bot(\chi)$ -- solid line
and $C_N(\chi)$ -- dashed line.}
\label{fig-Cppn}
\end{figure}
\begin{figure}
\epsfxsize=5.5in
\leavevmode
\epsffile[30 310 535 510]{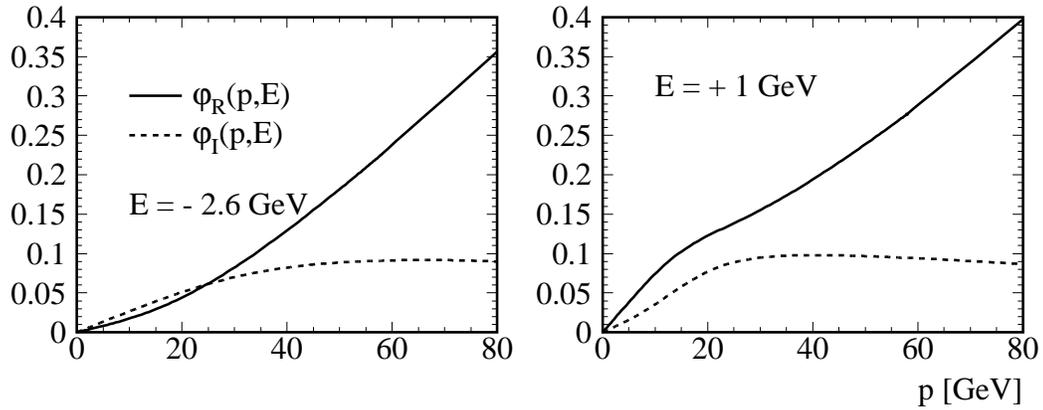}
\vskip-0.2cm
\caption{Momentum dependence of the functions $\varphi_R(p,E)$
(solid lines) and  $\varphi_I(p,E)$ (dashed lines):
a) $E$ = -2.6 GeV, and b) $E$ = 1 GeV.}
\label{fig-phiRI}
\end{figure}
The energy dependence of all the coefficient functions
$C(\chi)$ is very weak and can be neglected.
In Fig.\ref{fig-Cppn}a the coefficient functions $C^0_{^\|}(\chi)$
and $C^1_{^\|}(\chi)$ are shown. It is evident that for maximal
and minimal values of $\chi=\pm1$ the top quark is nearly maximally
polarized along the direction of the incoming electron.
This demonstrates that polarization studies close to threshold
are very promissing indeed. The other components of the top
polarization can be also interesting and the corresponding
coefficient functions are plotted in Fig.\ref{fig-Cppn}b.
Momentum dependence of the functions $\varphi_R(p,E)$
and  $\varphi_I(p,E)$ is shown in Fig.\ref{fig-phiRI}
for two energies in the threshold region.

\section{Semileptonic decays of heavy quarks}
The energy and angular distributions of the charged leptons and
the neutrinos are sensitive to the polarization of the decaying
heavy quark. Therefore they can be used in determination of
this polarization. Furthermore the basic assumption about the
V-A Lorentz structure of the charged weak current can be tested.
In \cite{CJ} compact analytic formulae have been obtained
for the distributions of the charged lepton and the neutrino.
These formulae  agree with the energy spectra which have been
obtained in \cite{JK89b}   and also
with the results of \cite{CJK91} and \cite{CJKK}
for the joint angular and energy distribution
of the charged lepton in top, charm and bottom quark
decays. The QCD corrected triple differential distribution
of the charged lepton  for the semileptonic decay  of the polarized
quark with the weak isospin $I_3=\pm 1/2$
can be written in the following way \cite{CJ}:
\begin{eqnarray}
{{\rm d}\Gamma^{\pm} \over {\rm d}x\,{\rm d}y\,{\rm d}\cos\theta  }
&& \sim\qquad \left[\,
{\rm F}^\pm_0(x,y) + {\cal P}\cos\theta\,{\rm J}^\pm_0(x,y)\,
\right]
\nonumber\\
&& -\; {2\alpha_s\over3\pi}\;
  \left[\,
  {\rm F}^\pm_1(x,y) + {\cal P}\cos\theta\,{\rm J}^\pm_1(x,y)\,
  \right]
\nonumber\\
\label{eq-dGdxdydth}
\end{eqnarray}
In the rest frame of the decaying heavy quark
$\theta$ denotes the angle between the
polarization vector $\vec{\cal P}$ of the heavy quark and the
direction of the charged lepton,
${\cal P}=|\,\vec{\cal P}\,|$,
$x= 2Q\ell/Q^2$ and $y= 2\ell\nu/Q^2$ where
$Q$, $\ell$ and $\nu$ denote the four-momenta of the decaying
quark, charged lepton and neutrino. Eq.(\ref{eq-dGdxdydth})
describes also the triple differential
distribution of the neutrino for  $I_3=\mp 1/2$. In this case,
however, $x= 2Q\nu/Q^2$  and
$\theta$ denotes the angle between
$\vec{\cal P}$ and the three-momentum of the neutrino.
The functions ${\rm F}^\pm_0(x,y)$ and ${\rm J}^\pm_0(x,y)$
corresponding to Born approximation read:
\begin{eqnarray}
{\rm F}^+_0(x,y) &=& x (x_m-x)
\label{eq-F0p}\\
{\rm J}^+_0(x,y) &=& {\rm F}^+_0(x,y)
\label{eq-J0p}\\
{\rm F}^-_0(x,y) &=& (x-y) (x_m-x+y)
\label{eq-F0m}\\
{\rm J}^-_0(x,y) &=& (x-y) (x_m-x+y-2y/x)
\label{eq-J0m}
\end{eqnarray}
where $x_m=1-\epsilon^2$, $\epsilon^2= q^2/Q^2$, and $q$
denotes the four-momentum of the quark originating from
the decay.
The functions ${\rm F}^\pm_1(x,y)$ and
${\rm J}^+_1(x,y)$ correspond to
the first order QCD corrections and
are given in \cite{CJ}.
Eq.(\ref{eq-J0p}) implies that for the top and charm quarks
the double differential angular-energy distribution
of the charged lepton is the product of the energy
distribution and the angular distribution.
QCD corrections essentially do not spoil this factorization
\cite{CJK91}. For the neutrino such factorization
does not hold, c.f. eqs.(\ref{eq-F0m}) and (\ref{eq-J0m}).
After integration over $x_\nu$ the angular dependence
of the neutrino distribution is much weaker than for the charged
lepton. For the bottom quark the roles of the charged lepton
and the neutrino are reversed. In the following part of this section
we limit our discussion to the semileptonic decays of the top quark.
Semileptonic decays of charm and bottom quarks will be considered
in the subsequent section.
For the top quark the decay rate is dominated
by the mode $t\to bW^+$, so neglecting the width of $W$ one fixes
$y$ in  eq.(\ref{eq-dGdxdydth}) at the value $y= m_{\rm w}^2/m_t^2$.

\begin{table}
\caption{Angular dependence of the distributions of $W$ bosons,
neutrinos and less energetic leptons in  $t\to bW\to be^+\nu$
or light quark jets in $t\to bW\to b\bar du$ decays.}
\label{tab:ang}
\begin{tabular*}{\textwidth}{@{}c@{\extracolsep{\fill}}crrr}
\hline
  &            & $m_t$=150  &  $m_t$=175 &  $m_t$=200  \\
\hline   \\
$h_\nu(y)$    & $1- {12y(1-y+y\ln y)\over(1-y)^2(1+2y)}$&
  -0.521      &  -0.311   &    -0.127    \\      \\
$h_{\rm w}(y)$& ${1-2y\over 1+2y}$   &
   0.275      &   0.410   &     0.515    \\      \\
$h_<(y)$& $1- {6y\{1-y-2y\ln[(1+y)/(2y)]\}\over(1-y)^2(1+2y)}$&
   0.464     &   0.509   &     0.559    \\
\\
\hline
\end{tabular*}
\end{table}
In the rest frame of the decaying $t$ quark the angular distributions
of the decay products are sensitive to its polarization.
Let us define
the angle $\theta_{\rm w}$ between $W$ boson three-momentum and
the polarization three-vector $\vec{\cal P}$.
Note that ${\cal P}= |\vec{\cal P}\,| = 1$ corresponds to fully
polarized and ${\cal P}=0$ to unpolarized top quarks. We define also
the angles $\theta_+$ and $\theta_0$ between $\vec{\cal P}$ and the
directions of the charged lepton and the neutrino, respectively,
and $\theta_<$ for the less energetic lepton in semileptonic
or less energetic light quark in hadronic decays.
For the sake of simplicity let us
confine our discussion to Born approximation
and consider semileptonic $t\to bW\to b\ell^+\nu$
and hadronic $t\to bW\to b\bar d u$ decays.\\
The angular distribution of the charged lepton
is of the form
\begin{equation}
{ {\rm d}N\over{\rm d}\cos\theta_+} =
{1\over 2}\, \left[\, 1\,+
\,{\cal P}\cos\theta_+ \right]
\label{eq:elec1}
\end{equation}
which follows from the
factorization of the angular-energy distribution
into the energy and angular dependent parts.
This factorization holds for arbitrary top mass below
and above the threshold for decays into real $W$ bosons
\cite{JK89b}. It is noteworthy that for ${\cal P}$=1 the angular
dependence in (\ref{eq:elec1}) is maximal because
a larger coeffecient multiplying $\cos\theta_+$ would be
in conflict with positivity of the decay rate.
Thus the polarization analysing power of
the charged lepton angular distribution
is maximal and hence far superior to other
distributions discussed in the following. In particular
the angular distribution of the neutrino reads \cite{JK94}:
\begin{equation}
{ {\rm d}N\over{\rm d}\cos\theta_0} =
{1\over 2}\, \left[\, 1\,+
\,h_\nu(y){\cal P}\cos\theta_0 \right]
\label{eq:neut1}
\end{equation}
where  $h_\nu(y)$ is given in Table \ref{tab:ang}. The
distribution of the direction of $W$ can be easily obtained.
Only the amplitudes for the helicity states of $W$
$\lambda_{\rm w} = -1$ and $\lambda_{\rm w} = 0$
are allowed and their contributions to the decay
rate are in the ratio $\;2y\; :\; 1\;$ \cite{GilKau}.
The corresponding angular distributions are of the form
\begin{equation}
{{\rm d}N_{-1,0}\over{\rm d}\cos\theta_{\rm w}} = {1\over2}
\left(1\mp {\cal P} \cos\theta_{\rm w} \right)
\end{equation}
After summation over the $W$ polarizations
the following angular dependence is obtained:
\begin{equation}
{{\rm d}N\over{\rm d}\cos\theta_{\rm w}} = {1\over2}
\left[1 + h_{\rm w}(y) {\cal P} \cos\theta_{\rm w} \right]
\end{equation}
where  $h_{\rm w}(y)$ is also given in Table \ref{tab:ang}.
It is evident that the charged lepton
angular distribution is significantly
more sensitive towards the polarization of $t$ than the
angular distributions of $W$ and $\nu$.
The charged lepton is likely to be the less energetic
lepton because its energy spectrum is softer
than that of the neutrino.
For large values of $m_t$ the angular distribution
of the less energetic lepton
\begin{equation}
{{\rm d}N\over{\rm d}\cos\theta_<} = {1\over2}
\left[1 + h_<(y) {\cal P} \cos\theta_< \right]
\end{equation}
is a more efficient analyser of top polarization than the angular
distribution of neutrinos. For $m_t$ in the range 150-200 GeV
it is also better than the direction of $W$,
c.f. Table \ref{tab:ang}.

\begin{figure}
\epsfxsize=5.5in
\leavevmode
\epsffile[60 300 490 525]{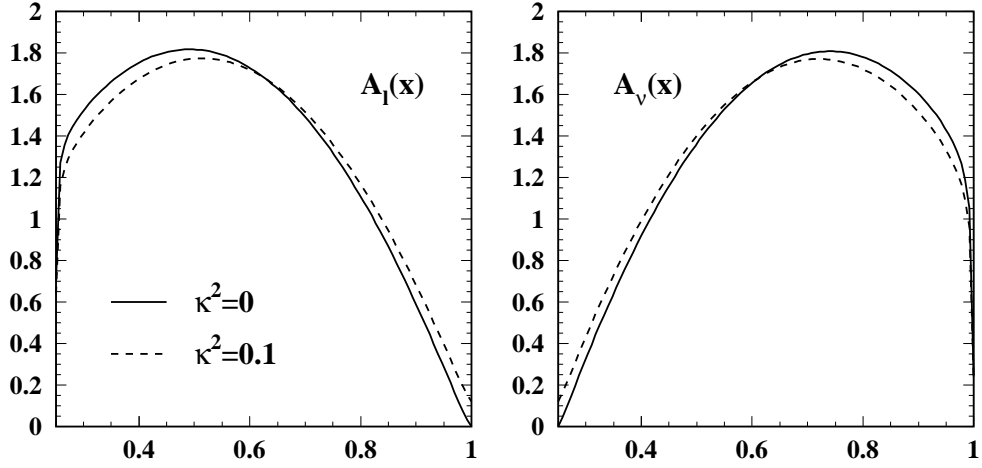}
\vskip-0.5cm
\caption{Energy distributions a) ${\rm A_l}(x_\ell)$ of the charged
lepton and b) ${\rm A}_\nu(x_\nu)$ of the neutrino
for  the standard model V-A coupling ($\kappa^2=0$)
and an  admixture of V+A current
($\kappa^2=$0.1) for $y=$0.25 and $\alpha_s=$0.11. }
\label{fig-Ael}
\end{figure}

The normalized distributions of leptons
including first order QCD corrections
can be cast into the following form:
\begin{eqnarray}
{ {\rm d}N\over{\rm d}x_\ell\,{\rm d}\cos\theta_+} &=&
{1\over 2}\,\left[\,
{\rm A_l}(x_\ell)\, +\, {\cal P}\cos\theta_+\,{\rm B_l}(x_\ell)\,
\right]
\nonumber\\
&& \\
{ {\rm d}N\over{\rm d}x_\nu\,{\rm d}\cos\theta_0} &=&
{1\over 2}\,\left[\,
{\rm A}_\nu(x_\nu)\, +\, {\cal P}
\cos\theta_0\,{\rm B}_\nu(x_\nu)\,\right]
\nonumber\\
\end{eqnarray}
Assuming the Standard Model
V-A structure of the charged current
the spectrum of the charged lepton vanishes at $x_\ell=1$
and the spectrum of the neutrino does not vanish at $x_\nu=1$.
The latter spectrum is also
significantly harder, see solid lines in Fig.\ref{fig-Ael}a-b.
\begin{figure}
\epsfxsize=5.5in
\leavevmode
\epsffile[60 300 490 525]{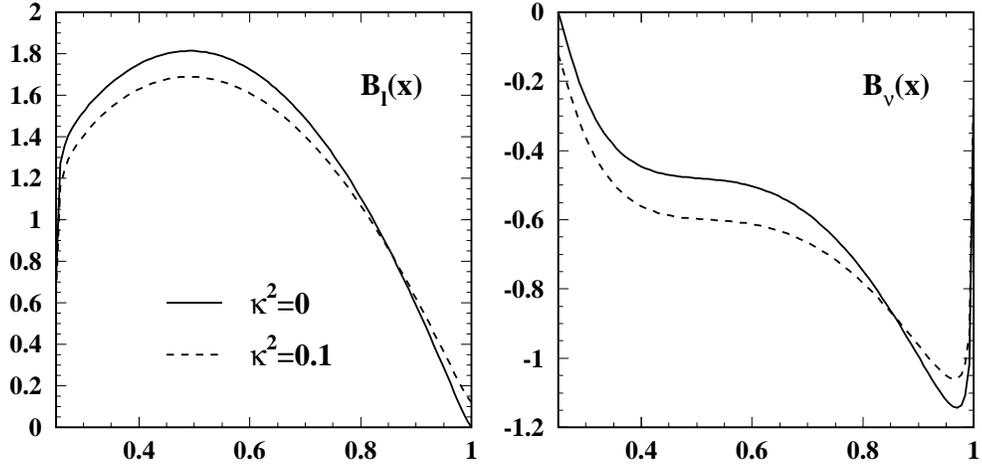}
\vskip-0.5cm
\caption{Angular-energy distribution functions
in the Standard Model ($\kappa^2=0$)
and for the admixture of V+A current ($\kappa^2=$0.1):
a) ${\rm B_l}(x_\ell)$ for the charged
lepton and b) ${\rm B}_\nu(x_\nu)$ for the neutrino,
$y$=0.25 and $\alpha_s$=0.11.}
\label{fig:Bel}
\end{figure}
For V+A coupling the charged lepton and the neutrino energy
spectra would be interchanged in comparison to the V-A case.
In \cite{JK94} effects have been studied
of a small admixture of non-standard V+A current on
distributions of leptons. The $tbW$ vertex has been parametrized
as
\begin{eqnarray}
g_{_V}\,\gamma^\mu\; +\; g_{_A}\,\gamma^\mu\gamma_5
\label{eq:VA}
\end{eqnarray}
where
$g_{_V} = ( 1+ \kappa)/\sqrt{1+\kappa^2}$
and $g_{_A} = ( -1+ \kappa)/\sqrt{1+\kappa^2}$
Hence $\kappa=0$ corresponds to pure V-A and $\kappa=\infty$
to V+A. In Fig. \ref{fig-Ael}a-b the lepton
spectra are plotted corresponding to $\kappa^2=0.1$, see
dashed lines. It can be seen that
the deviations from the results of the Standard Model (solid lines)
are rather small.
In Fig.\ref{fig:Bel} the functions
${\rm B_l}(x)$  and    ${\rm B}_\nu(x)$
are shown as solid lines for $y=$0.25 and $\alpha_s=$0.11 \cite{JK94}.
The effect of non-standard coupling defined in eq.(\ref{eq:VA})
is much stronger for the polarization dependent distribution
of neutrinos, see dashed lines in Fig.\ref{fig:Bel} corresponding
to $\kappa^2=0.1$ \cite{JK94}.

\section{Polarized bottom and charm quarks}

Polarization studies for heavy flavors at LEP\cite{Mele,Roudeau}
and SLC\cite{Abe95} are a new and interesting field
of potentially fundamental significance.
According to the Standard Model $Z^0\to b\bar b$ and
$Z^0\to c\bar c$ decays
can be viewed as sources of highly polarized heavy quarks.
The degree of longitudinal polarization is fairly large,
amounting to $\langle P_b\rangle = -0.94$ for $b$
and $\langle P_c\rangle = -0.68$ for $c$ quarks~\cite{Kuehn1}.
The polarizations depend weakly on the production angle.
QCD corrections to Born result are about 3\% \cite{KPT}.
Therefore there is no doubt that the heavy quarks produced
at the $Z^0$ resonance are polarized. However, this prediction
still awaits a firm experimental verification. Unfortunately,
these are hadrons rather than quarks which are registered
in the detectors and the quantitative theoretical description
of the spin transfer during the time development of a heavy quark jet
is still lacking. Thus, it is not clear in which way the original
high degree of polarization is reflected in the properties of
jets containing heavy flavours.
It has been proposed\cite{Nacht,Efrem} that non-zero helicities
and chiralities of heavy quarks may result in non-zero values
of two-particle momentum correlations for the most energetic
particles in jets:
\begin{eqnarray}
\Omega_{hel} = {\bf t\cdot\left( k_1\times k_2 \right)}
\qquad {\rm and} \qquad
\Omega_{chi} = {\bf t\cdot\left( k_+\times k_- \right)}
\nonumber
\end{eqnarray}
where for $\Omega_{chi}$ only particles of opposite electric
charges are considered. However, in \cite{Abe95} a negative result
has been recently reported of the search for the asymmetries
in distributions of $\Omega_{hel}$ and $\Omega_{chi}$
No definite conclusion follows from this finding because
no detailed theory exist relating these corelation functions
with the primordial polarizations of the heavy quarks.

It seems more interesting to look for some signatures
of the primordial polarization in those processes
for which theoretical description is more reliable.
Semileptonic decays of heavy flavors belong to this cathegory.
Recently there has been considerable progress in the theory of
the inclusive semileptonic decays
of heavy flavor hadrons\cite{BigiCEC}.
In the framework of Heavy Quark Effective Theory (HQET) and $1/m_Q$
expansion it has been shown that in the leading order
the lepton spectra for the decays of hadrons
coincide with those for the decays of
free heavy quarks \cite{CGG}.
Away from the endpoint region
there are no $\Lambda_{QCD}/m_Q$
corrections to this result \cite{CGG}
and $\Lambda^2_{QCD}/m_Q^2$ corrections have been
calculated in \cite{Bigi,wise} for $B$ mesons and
in \cite{wise} for polarized $\Lambda_b$ and $\Lambda_c$ baryons.
For some decays the results are similar to those of the well-known
ACCMM model \cite{altar}.
The corrections to charm decays are larger than for bottom
and convergence of
$1/m_Q$  expansion is poorer \cite{Shifman}.
In \cite{CJKK} order $\alpha_s$ perturbative QCD
corrections have been calculated to the angular and energy
distributions of leptons in semileptonic decays of polarized
charm and bottom quarks.
Thus a complete theoretical description exists of
the inclusive lepton spectra which should be accurate
up to the level of few percent.
Moreover, it has been pointed out \cite{CJKK,BR}
that for semileptonic chanels not only the charged leptons
but also the neutrinos can be registered as a missing
energy-momentum. In consequence the sensitivity to the primordial
polarization can be increased and simultaneously
ambiguities in the process of jet fragmentation
can be significantly reduced\cite{BR}.
The real drawback is that due to hadronization the net longitudinal
polarization of the decaying  $b$ and $c$ quarks is drastically
decreased. In particular these $b$ quarks become depolarized
which are bound in $B$ mesons
both produced directly and from $B^*\to B\gamma$
transitions\footnote{$B^*$ and $D^*$ mesons
from fragmentations of polarized $b$ and $c$ quarks retain some
information on the primordial polarization. It is plausible that
a quark with helicity -1/2 fragments into a state
of helicity -1 more frequently than into that of helicity +1.
In electromagnetic transitions, however this information is
lost unless the polarization of real or virtual $\gamma$
is measured. $D^*\to D\pi$ transitions might be more useful
in this respect.}.
The signal is therefore significantly reduced.
Only those $b$'s (a few percent) which fragment directly
into $\Lambda_b$  baryons retain information on the original
polarization \cite{Bjo}.
Polarization transfer from a heavy
quark $Q$ to the corresponding $\Lambda_Q$ baryon is 100\%
\cite{CKPS} at least in the limit $m_Q\rightarrow\infty$.
Thus, a large net polarization is expected for heavy quarks
in samples enriched with heavy $\Lambda_b$ and $\Lambda_c$ baryons.
Since semileptonic
decays are under control it is possible to measure these
polarizations. Many new opportunities arise, polarization
studies for other decay chanels among them. One of the most
interesting may be studies of non-perturbative
effects in fragmentation of bottom and charm quarks.
Comparison of polarizations for $\Lambda_b$ and $\Lambda_c$
baryons can be instrumental in studying non-perturbative
corrections to the spin transfer in fragmentation.
This will be possible only if experimentalists can separate
directly produced baryons from those from resonances.
Assuming that this is possible and anticipating further progress
in HQET as well as in perturbative QCD calculations one may expect
that polarization studies for $b$ systems at LEP
will offer new opportunities to test the Standard Model.
Recently, the ALEPH collaboration reported a preliminary
result on $\Lambda_b$ polarization
$P_{\Lambda_b} = - 0.30 ^{+ .32}_{- .27}\pm .04$ \cite{Roudeau}.
This result which is well below theoretical expectations
indicates that the sample may be contaminated
with $\Lambda_b$'s from decays of other beautiful baryons.

\section{Acknowledgements}
I would like to dedicate this work to Professor Kacper
Zalewski on the occasion on his sixtieth birthday.
Many of us owe Professor Zalewski a great debt for
the stimulus and friendship he has given us over
many years. I am particularly indebted to him
for a help at a crucial stage in my academic life.
I hope that this meeting has helped in some way
to express that gratitude as well as our best wishes
for the future.

\end{document}